\documentclass[runningheads]{llncs}

\usepackage[utf8]{inputenc}
\usepackage[T1]{fontenc}
\usepackage{mathptmx}
\usepackage{soul}\setuldepth{article}
\usepackage{cite}
\usepackage{graphicx}
\usepackage{url}
\usepackage{makecell}
\def\hb{\hbox to 10.7 cm{}}
\usepackage{fancyhdr}
\usepackage{iftex}

\ifTUTeX
\usepackage{fontspec}
\else
\usepackage[T1]{fontenc}
\usepackage[utf8]{inputenc} 
\DeclareUnicodeCharacter{200B}{{\hskip 0pt}}
\fi

\begin{document}

\pagestyle{headings}
\def\thepage{}

\title{Towards Grad-CAM Based Explainability in a Legal Text Processing Pipeline}
%
%
\author{Łukasz Górski\inst{1}\orcidID{0000-0003-0871-6575} \and\\
Shashishekar Ramakrishna\inst{2,3}\orcidID{0000-0002-2832-0415} \and\\
Jędrzej M. Nowosielski\inst{1}\orcidID{0000-0002-9627-3008}}
\authorrunning{Ł. Górski et al.}
%
\institute{Interdisciplinary Centre for Mathematical and Computational Modelling, Univ. of Warsaw \and
	Freie Universit{\"a}t Berlin, Berlin\and
	EY - AI Labs, Bangalore}
\maketitle              

\fancypagestyle{empty}{%
	\fancyhf{}
	\fancyfoot[L]{\footnotesize{Presented at the Workshop on EXplainable \& Responsible AI in Law (XAILA) \\ at 33rd International Conference on Legal Knowledge and Information Systems​ (JURIX)\\ 9 - 11 December, 2020}}
	\renewcommand{\footrulewidth}{0.6pt}
	\renewcommand{\headrulewidth}{0pt}
}

\begin{abstract}
Explainable AI (XAI) is a domain focused on providing interpretability and explainability of a decision-making process. In the domain of law, in addition to system and data transparency, it also requires the (legal-) decision-model transparency and the ability to understand the model's inner working when arriving at the decision. This paper provides the first approaches to using a popular image processing technique, Grad-CAM, to showcase the explainability concept for legal texts. With the help of adapted Grad-CAM metrics, we show the interplay between the choice of embeddings, its consideration of contextual information, and their effect on downstream processing.

\end{abstract}

\section{Introduction}
\label{section:Introduction}
 Advancements in the domain of AI and Law have brought additional considerations regarding models development, deployment, updating and their interpretability. This can be seen with the advent of machine-learning-based methods, which naturally exhibit a lower degree of explainability than traditional knowledge-based systems. Yet, knowledge representation frameworks that handle legal information, irrespective of their origin, should cover the pragmatics or context around a given concept and this functionality should be easily demonstrable.

Explainable AI (XAI), is a domain which has focused on providing interpretability and explainability to a decision making process. In the domain of law, interpretability and explainability are more than dealing with information/data transparency or system transparency~\cite{bibal2020legal} (henceforth referred to as \textit{ontological view}). It additionally requires the (legal-) decision-model transparency, the ability to understand the model's inner working when arriving at the decision (\textit{epistemic view}). In this paper, we aim to present the system's user and architect with a  set of tools that facilitate the discovery of inputs that contribute to convolutional neural network's (CNN's) output to the greatest degree, by adapting the Grad-CAM method, which originated from the field of computer vision. We adapt this method to the legal domain and show how it can be used to achieve a better understanding of a given system's state and explain how different embeddings contribute to end result as well as to optimize this system's inner workings. While this work is concerned with the ontological perspective, we aim this as a stepping stone for another related perspective, where the legally-based positions are connected with explanation thus providing the ability to explain the decisions to its addressee. This paper addresses mainly the technical aspects, showing how Grad-CAMs can be applied to the legal texts, describing the text processing pipeline - taking this as a departing point for deeper analyses in future work. We aim to present this technical implementation as well as the quantitative comparison metrics as the main contribution of the paper.

The paper is structured as follows. State-of-the-art is described in Section~\ref{section:Related Work}. Section \ref{section:Methodology} describes the methodology, which includes the metrics used for results quantification.  The architecture used for experiments is described in Section \ref{section:Architecture}. Section~\ref{section:Experiments} talks about the different datasets used and the experimental setup. The outcomes are described in Section \ref{section:Results}. Finally, Section \ref{section:Conclusion} provides a conclusion and future work.

%
%

\section{Related Work}
\label{section:Related Work}

The feasibility of using different - contextual (e.g. BERT) and non-contextual (e.g. word2vec) - embeddings was already studied outside the domain of law. In~\cite{arora2020contextual}, it was found that the usage of more sophisticated, context-aware methods is unnecessary in the domains where labelled data and simple language are present. As far as the area of law is concerned, the feasibility of using the domain-specific vs. general embeddings (based on word2vec) for the representation of Japanese legal texts was investigated, with the conclusion that general embeddings have an upper hand~\cite{Dtang2019examination}. The feasibility of using BERT in the domain of law was also already put under scrutiny as well. In~\cite{condevaux2019weakly} its generic pretrained version was used for embeddings generation and it was found that large computational requirements may be a limiting factor for domain-specific embedding creation. The same paper concluded that the performance of the generic version is lower when compared with law-based non-contextual embeddings. On the other hand, in \cite{rossi2019legal}, BERT versions trained on legal judgments corpus (of 18000 documents) were used and it was found that training on in-domain corpus does not necessarily offer better performance compared to generic embeddings. In \cite{elwany2019bert} contradictory conclusions were reached: the system's performance significantly improves when using pre-trained BERT on a legal corpus. Those results suggest that introduction of XAI-based methods might be a \textit{condition sine qua non} for a proper understanding of general language embeddings and their feasibility in the domain. 

Grad-CAMs are explainability method originating from computer vision~\cite{selvaraju2017grad}. It is a well established post-hoc explainability technique when CNNs are concerned. Moreover, Grad-CAM method passed independent sanity checks~\cite{NEURIPS2018_294a8ed2}. Whilst it is mainly connected with the explanations of deep learning networks used with image data, it has already been adapted for other areas of application. In particular, CNN architecture for text classification was described in~\cite{kim2014convolutional}, and there exists at least one implementation which extends this work with Grad-CAM support for explainability~\cite{haebinshi_git}. Grad-CAMs were already used in the NLP domain, for (non-legal) document retrieval~\cite{choi2020interpreting}. Herein we build upon this work and investigate the feasibility of using this method for the legal domain, in particular allowing for the visualisation of context-dependency of various word embeddings. Legal language is a special register of everyday language and deservers investigation on its own. The evolution of legal vocabulary can be precisely traced to particular statutes and precedential judgments, where it is refined and its boundaries are tested~\cite{rissland2003ai}. Many terms have thus a particular legal meaning and efficacy and tools that can safeguard final black-box models' adherence to the particularities of legal language are valuable.

The endeavours aimed at using XAI methods in the legal domain, similar to this paper, have already been undertaken recently. In~\cite{branting2020scalable} an Attention Network was used for legal decision prediction - coupling it with attention-weight-based text highlighting of salient case text (though this approach was found to be lacking). The possibility of explaining the BERT's inner workings was already investigated by other authors, and it was already subject to static as well as dynamic analyses. An interactive tool for the visualisation of its learning process was implemented in~\cite{hoover2019exbert}. Machine-learning-based evaluation of context importance was performed in~\cite{savelka2019improving}; therein it was found that accounting for the content of a sentence's context greatly improves the performance of legal information retrieval system.

However, the results mentioned hereinbefore do not allow for direct and easily interpretable comparison of different types of embeddings and we aim to explore an easy plug-in solution facilitating this aim.

\section{Methodology}
\label{section:Methodology}
We study the interplay between the choice of embeddings, its consideration of contextual information, and its effect on downstream processing. For this work, a pipeline for comparison was prepared, with the main module being the embedder, classification CNN and metric-based evaluator. All the parts are easily pluggable, allowing for extendibility and further testing of a different combinations of modules.

The CNN used in the pipeline was trained for classification. We use two different datasets for CNN training (as well as testing)~\footnote{Section~\ref{section:Dataset}, provides a detailed discussion on the considered datasets}:

\begin{enumerate}
\item The Post-Traumatic Stress Disorder (PTSD)~\cite{Moshiashwili2015veterans} dataset~\cite{walker2019automatic},  where rhetorical roles of sentences are classified.
\item Statutory Interpretation - Identifying Particular (SIIP) dataset~\cite{savelka2019improving}, where the sentences are classified into four categories according to their usefulness for a legal provision's interpretation. 
\end{enumerate}

Whilst many methods have already been used for the analysis of aforementioned datasets (including regular expressions, Naive Bayes, Logistic Regression, SVMs~\cite{walker2019automatic}, or Bi-LSTMs~\cite{ahmad2020understanding}), we are unaware of papers that use (explainable) CNNs for this tasks. On the other hand, usage of said CNN should not be treated as the main contribution of this paper, as the classification network is treated only as an exemplary application, warranting conclusions regarding the paper's main contribution, i.e. the context-awareness of various embeddings when used in the legal domain. 

Further down the line, the embeddings are used to transform CNN input sentences into vectors, with vector representation for each word in a sentence concatenated. Herein our implementation is based on the prior work~\cite{haebinshi_git}\cite{kim2014convolutional}. 

\subsection{Comparison metrics}
\label{subsection:comparison-metrics}
Grad-CAM heatmaps are inherently visual tools for data analysis. In computer vision, they are commonly used for qualitative determination of input image regions that contribute to the final prediction of the CNN. While they are an attractive tool for a qualitative analysis of a single entity, they should be supplemented with other tools for easy comparison of multiple embeddings~\cite{krakov2013comparing} and to facilitate quantitative analysis.  Herein  the following metrics are introduced and adapted to the legal domain:

\begin{enumerate}
	\item Fraction of elements above relative threshold $t$ ($\mathcal{F}(v, t)$) 
	\item Intersection over union with relative thresholds $t_1$ and $t_2$ ($\mathcal{I}(v_1, v_2, t_1, t_2)$)
\end{enumerate}

The first metric, $\mathcal{F}(t)$, is designed to measure the CNN network attention spread over words present in the given input, i.e, what portion of the input is taken into account by CNN in the case of a particular prediction. It is defined as a number of elements in a vector that are larger than the relative threshold $t$ multiplied by the maximum vector value divided by the length of this vector.

The second metric, $\mathcal{I}(v_1, v_2, t_1, t_2)$, helps to compare two predictions of two different models given the same input sentence. It answers the question of whether two models, when given the same input sentence, `pay attention' to the same or different chunk(s) of the input sentence. It takes as arguments two Grad-CAM heatmaps ($v_1$ and $v_2$), binarizes them using relative thresholds ($t_1$ and $t_2$) and finally calculates standard intersection over union. It quantifies the relative overlap of words considered important for the prediction by each of two models.   

\section{System Architecture}
\label{section:Architecture}

\begin{figure}[htb]
\includegraphics[width=11cm]{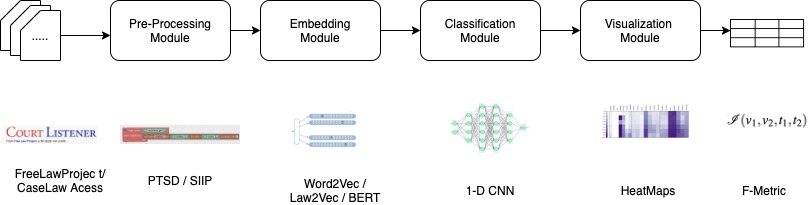}
\caption{System Architecture }
\label{fig:architecture}
\end{figure}

The architecture, as shown in Fig~\ref{fig:architecture}, is designed to implement the methodology described in~\ref{section:Methodology} and comprises  four main modules, i.e.: preprocessing module, embedding module, classification module and visualization module. The pre-processing module uses some industry \textit{de facto} standard text processing libraries for spelling correction, sentence detection, irregular character removal, etc.  The embedding module houses a plug-in system to handle different variants of embeddings, in particular BERT and word2vec. The classification module houses simple 1D CNN which facilitates explainability method common in computer vision i.e. Grad-CAM. The visualization module is used for heatmap generation and metric computation.

The output from the pre-processing module is fed into the embeddings module. The embeddings used are based on variants of BERT and word2vec. In addition to the pre-trained ones, raw data from CourtListener~\cite{courtlistner2020} dataset was used for training embeddings creation. 

Within the frame of the classification module, the output from the embeddings module is fed into a 1D convolutional layer followed by an average pooling layer and fully-connected layers with dropout and softmax~\cite{kim2014convolutional}. Although CNN architectures stem from computer vision where an image forms the input of the network, the use of CNN for the sequence of word vectors as an input is reasonable. In a sentence relative positions of words convey meaning. It is similar to an image where relative positions of pixels convey information, with the difference being about dimensionality. Standard image is 2D while a sentence is a 1D sequence of words, therefore we use the 1D CNN for the task of sentence classification. 

With Grad-CAM technique it is possible to produce a class activation map (heatmap) for a given input sentence and predicted class. Each element of the class activation map corresponds to one token and indicates its importance  in terms of the score of the particular (usually the predicted) class. The class activation map gives information on how strongly the particular tokens present in the input sentence influence the prediction of the CNN.   

The software stack used for the development of this system was instrumented under Anaconda 4.8.3 (with Python 3.8.3). Tensorflow v. 2.2.0 was used for CNN instrumentation and Grad-CAMs calculations (with the code itself expanding prior implementation available at~\cite{haebinshi_git}). Spacy 2.1.8 and blackstone 0.1.15 were used for CourtListener text cleaning. Various BERT implementations and supporting codes  were sourced from Huggingface libraries: transformers v. 3.1.0, tokenizers v. 0.8.1rc2, nlp v. 0.4.0.
Two computing systems available at ICM University of Warsaw were exploited for the experiments. Text cleaning was performed using the okeanos system (Cray XC40) and main calculations were run on rysy GPU cluster (4x Nvidia Tesla V100 32GB GPUs).
\section{Experiments}
\label{section:Experiments}

\subsection{Datasets}
\label{section:Dataset}

As stated in Section~\ref{section:Methodology}, we use two different datasets for experiments. The PTSD dataset is from the U.S. Board of Veterans' Appeals (BVA) from 2013 through 2017. The dataset deals with the decisions from adjudicated disability claims by veterans for service-related post-traumatic stress disorder (PTSD)~\cite{Moshiashwili2015veterans}. The dataset itself is well-known and has already been studied by other authors. It annotates a set of sentences originating from 50 decisions issued by the Board according to their function in the decision~\cite{walker2019automatic} \cite{walker2017SemanticTypes} \cite{Savelka2017SentenceBD}. The classification consists of six elements: \textit{Finding Sentence}, \textit{Evidence Sentence}, \textit{Reasoning Sentence}, \textit{Legal-Rule Sentence}, \textit{Citation Sentence}, \textit{Other Sentence}. 

The SIIP dataset pertains to the United States Code 5 § 552a(a)(4) provision and aims to annotate the judgments that are most useful for interpretation of said provision. The seed information for annotation is collected from the court decisions retrieved from the Caselaw access project data. The sentences are classified into four categories according to their usefulness for the interpretation: \textit{High Value},  \textit{Certain Value}, \textit{Potential Value}, \textit{No Value} \cite{savelka2019improving}.

\subsection{Embeddings/Language Modeling}
We use pre-trained models as well as we train domain-specific models for the purpose of vector representation of texts. Many flavours of word2vec and BERT embedders were tested. The paper does not go into any details on the comparison of these pre-trained models (or other similar models) based on performance. This has been addressed in several other papers~\cite{martin2020camembert}\cite{sanh2019distilbert}\cite{hoover2019exbert}.

 For the word2vec a (slimmed down) GoogleNews model was used, with a vocabulary of $300 000$ words~\cite{eyealer_git}. In addition, Law2vec embeddings were also employed, which were trained on a large freely-available legal corpus, with 200 dimensions~\cite{law2vec}. For BERT, bert-base-uncased model was used, a transformer model consisting of 12 layers, 768 hidden units, 12 attention heads and 110M parameters. In addition to that, a slimmed-down version of BERT, DistilBERT was also tried, due to its accuracy being on the par with vanilla BERT, yet offering better performance and smaller memory footprint.

In addition to pretrained models, we have also tried training our own word2vec and BERT models. For this aim, a CourtListener~\cite{courtlistner2020} database was sourced. However, due to the large computational requirements of BERT training, a small subset of this dataset was chosen, consisting of 180MiB of judgments. Moreover, while several legal projects provide access to a vast database of US case-laws, it was found that the judgments available therein need to be further processed, as the available textual representations usually contain unnecessary elements, such as  page numbers or underscores, that hinder their machine processing. Our hand-written parser joined hyphenated words, removed page numbers and artifacts that were probably introduced by OCR-ing; furthermore, the text was split into sentences using spacy-based blackstone-parser. In line with other authors~\cite{choibankruptcy}, we have found it to be imperfect and failing in segmenting the sentences that contained period-delimited legal abbreviations (e.g. \textit{Fed.} - Federal). Thus it was supplemented with our own manually-curated list of abbreviations. The training was performed using DistilBERT model (for ca. 36 hours), as well as word2vec in two flavours, 200-dimensional (in line with the dimensionality of Law2Vec) and 768-dimensional (in line with BERT embeddings dimensionality).

As far as the BERT-based embeddings go, there is a number of ways in which they can be extracted from the model. One of the ways is taking embeddings for special \textit{CLS} token, which prefixes any sentence fed into BERT; another technique that was studied in the literature amounted to concatenating the model's final layer's values. The optimal technique is dependent on the task and the domain. Herein we have found the latter to offer better accuracy for downstream CNN training. The features for CNN processing consisted of tokenized sentences, together with embeddings for special BERT tokens (their absence would cause a slight drop in accuracy as well).


%

\section{Results}
\label{section:Results}

\subsection{Metric-based heatmap comparison}

%

\begin{table}[tbh]
	\centering
	\begin{tabular}{|l|c|c||c|c||c|c||}
		
		\hline 
		&
		\multicolumn{2}{c||}{$\mathcal{F}(0.15)$} & \multicolumn{2}{c||}{$\mathcal{F}(0.3)$} & \multicolumn{2}{c||}{$\mathcal{F}(0.5)$} \\ 
		\hline 
		& Mean & StdDev & Mean & StdDev & Mean & StdDev \\ 
		\hline 
		word2vec (GoogleNews) & 0.53 & 0.31 & 0.44 & 0.3 & 0.35 & 0.29 \\ 
		\hline
		Law2vec  & 0.6 & 0.3 & 0.52 & 0.32 & 0.42 & 0.33  \\
		\hline
		word2vec (CourtListener, 200d)  & 0.49 & 0.28 & 0.39 & 0.27 & 0.29 & 0.26 \\
		\hline
		word2vec (CourtListener, 768d)  & 0.48 & 0.28 & 0.38 & 0.28 & 0.29 & 0.27 \\
		\hline
		BERT(bert-base-uncased)  & 0.48 & 0.32 & 0.36 & 0.28 & 0.24 & 0.22 \\ 
		\hline 
		DistilBert(distilbert-base-uncased)  & 0.67 & 0.27 & 0.56 & 0.27 & 0.38 & 0.24 \\
		\hline
		DistilBERT(CourtListener)  & 0.47 & 0.39 & 0.47 & 0.39 & 0.44 & 0.39 \\
		\hline
	\end{tabular}
	\caption{Heatmap metric $\mathcal{F}$ for the PTSD dataset}
	\label{table:gradCAM_comparision_F} 
\end{table}

%

\begin{table}[tbh]
	\centering
	\begin{tabular}{|l|c|c||c|c||c|c||}
		\hline 
		& \multicolumn{2}{c||}{$\mathcal{I}(0.15)$} & \multicolumn{2}{c||}{$\mathcal{I}(0.3)$} & \multicolumn{2}{c||}{$\mathcal{I}(0.5)$} \\ 
		\hline 
		& Mean & StdDev & Mean & StdDev & Mean & StdDev\\ 
		\hline 
		\makecell[l]{word2vec (GoogleNews) --\\ \quad BERT(bert-base-uncased)} & 0.49 & 0.25 & 0.41 & 0.24 & 0.3 & 0.21 \\
		\hline
		\makecell[l]{Law2vec --\\ \quad BERT(bert-base-uncased)} & 0.51 & 0.26 & 0.43 & 0.25 & 0.34 & 0.25 \\
		\hline
		\makecell[l]{word2vec(CourtListener, 200d) --\\ \quad Law2Vec} & 0.65 & 0.25 & 0.58 & 0.27 & 0.51 & 0.31 \\
		\hline
		\makecell[l]{word2vec(CourtListener, 768d) --\\ \quad DistilBert(distilbert-base-uncased)} & 0.44 & 0.23 & 0. 35 & 0.22 & 0.26 & 0.21 \\
		\hline
	\end{tabular}
	\caption{Heatmap metric $\mathcal{I}$ for the selected pairs of embeddings for the PTSD dataset}
	\label{table:gradCAM_comparision_I} 

\end{table}

%

\begin{table}[tbh]
	\centering
	\begin{tabular}{|l|c|c|}
		\hline 
		& PTSD & SIIP \\ 
		\hline
		word2vec (GoogleNews) & 0.7 & 0.9\\
		\hline  
		Law2vec & 0.69 & 0.85  \\
		\hline
		word2vec (CourtListener, 200d) & 0.78 & 0.93 \\
		\hline
		word2vec (CourtListener, 768d) & 0.79 & 0.94 \\
		\hline
		BERT (bert-base-uncased) & 0.84 & 0.94  \\
		\hline
		DistilBERT (distilbert-base-uncased) & 0.85 & 0.94\\
		\hline
		DistilBERT (CourtListener) & 0.42 & 0.85 \\
		\hline
	\end{tabular}
	\caption{Test set accuracy.}
	\label{table:accuracy_training} 
\end{table}

\begin{figure}[htb]
	\includegraphics[width=11cm]{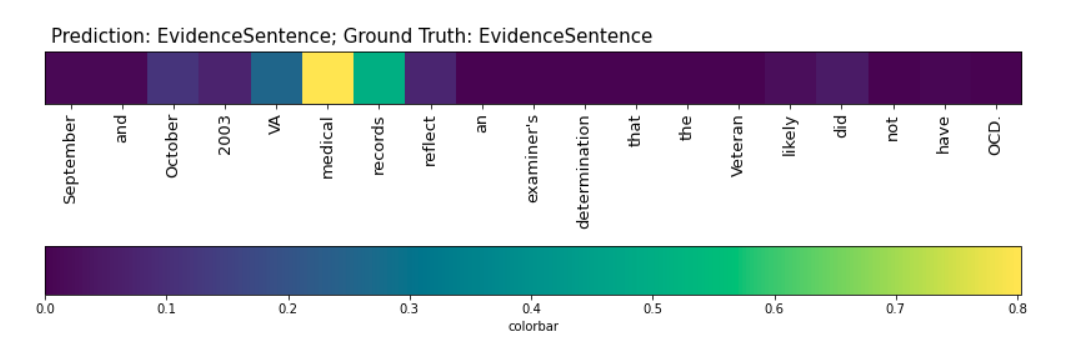}
	\caption{A sample heatmap for correct prediction with word2vec (CourtListener,768d) embedding}
	\label{fig:heatmap-Word2vec-CourtListener-768d-correct}
\end{figure}

\begin{figure}[htb]
\includegraphics[width=11cm]{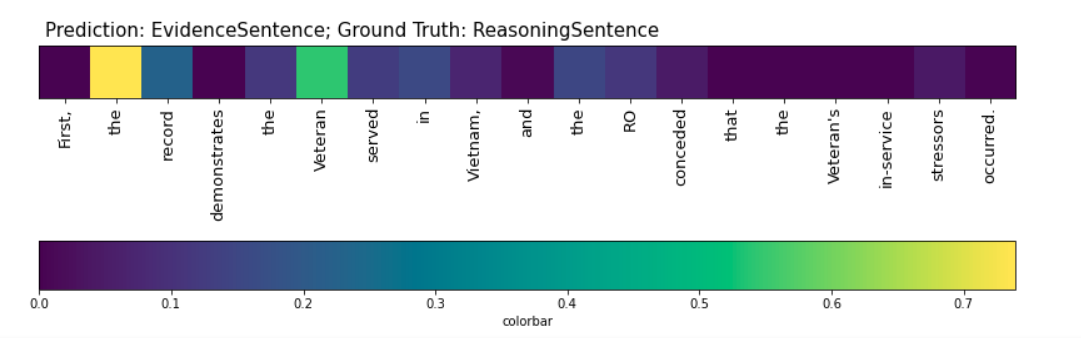}
\caption{A sample heatmap for failed prediction with word2vec (CourtListener,768d) embedding}
\label{fig:heatmap-Word2vec-CourtListener-768d-failed}
\end{figure}

A sample heatmap can be referenced in Fig.~\ref{fig:heatmap-Word2vec-CourtListener-768d-correct} and Fig.~\ref{fig:heatmap-Word2vec-CourtListener-768d-failed}, with a colorbar defining the mapping between the colors and values. Fig~\ref{fig:heatmap-Word2vec-CourtListener-768d-correct} clearly shows the area of CNN's attention, which can be quantified further down the line. This picture shows a properly classified sentence, a statement of evidence, defined by the PTSD dataset's authors as a description of a piece of evidence. CNN pays most attention to the phrase "medical records", which is in line with PTSD's authors' annotation protocols, where this kind of sentence describes a given piece of evidence (e.g. the records of testimony). We have found the sentence in Fig.~\ref{fig:heatmap-Word2vec-CourtListener-768d-failed} to be hard to classify for ourselves and it \textit{prima facie} seemed for us to be an example of evidentiary sentence. In the case of CNN, no distinctive activations can be spotted.

 Yet, we did not perform any detailed analyses of such images. Instead, we focus on two types of comparison using metrics defined in section~\ref{subsection:comparison-metrics}. The comparisons are designed to capture differences between different embeddings, particularly in terms of context handling. First, for a given embedding we calculate CNN network attention spread over words quantified  by metric $\mathcal{F}(t)$ averaged over all input sentences contained in the test set. Then we can compare the mean fraction of words (tokens) in the input sentences which contribute to prediction in the case of various embeddings. Criterion deciding if a particular word contributes to the p\textit{}rediction is, in fact, arbitrary and depends on class activation map (heatmap) binarization threshold. This is why we test a few thresholds, including $0.15$ as suggested in~\cite{selvaraju2017grad} for weakly supervised localization. Essentially high value of the fraction $\mathcal{F}(t)$ indicates that most word vectors in input sentence are taken into account by CNN during inference. Conversely, the low value of the fraction $\mathcal{F}(t)$ indicates that most word vectors in the input sentence are ignored by CNN during inference. The comparison results for the PTSD dataset are shown in Table~\ref{table:gradCAM_comparision_F} and Table~\ref{table:gradCAM_comparision_I} (SIIP dataset was omitted for brevity and due to the similarity with the presented PTSD dataset). The outstanding similarity between word2vec and Law2Vec can be spotted in Table~\ref{table:gradCAM_comparision_I}, due to both of those models belonging to the same class, as exhibited by the high value of $\mathcal{I}$ metric.


%
%
\subsection{Grad-CAM guided context extraction}
\label{section:grad-cam-guided-context-extraction}

The analysis of heatmaps and metrics presented hereinbefore proves that only a part of a given sentence contributes to a greater extent to final results. We have hypothesized that it is possible to decrease the amount of CNN's input data to those important parts without compromising the final prediction. In this respect, Grad-CAM was treated as a helpful heuristic that allows to identify the most important words for a given CNN in its training phase. For this experiment, the value of $\mathcal{F}$, for the threshold of $0.15$ was used to select a percentage of the most important words from a given training example. This in turn was used to compose a vocabulary (or white-list) of the most important words that were encountered during the training. Further down the line, this white-list was used during the inference and only the words present on the list were passed as input to the CNN. Nevertheless, the number of white-listed words allowed coherent sentences to be still passed into CNN (for example, the PTSD sentence 
\textit{However, this evidence does not make it clear and}, before white-listing amounted to \textit{However, this evidence does not make it clear and unmistakable.}). 

We have managed to keep accuracy up to the bar of an unmodified dataset using this procedure (e.g. 0.7 for PTSD-word2vec(GoogleNews) and 0.85 for PTSD-DistilBERT (distilbert-base-uncased). 

\section{Conclusion \& Future Work}
\label{section:Conclusion}
We presented the first approach to using a popular image processing technique, Grad-CAMs to showcase the explainability concept for legal texts. Few conclusions which we can be drawn from the presented methodology are: 
\begin{itemize}
	\item The mean value of $\mathcal{F}(t)$ is higher in the case of DistilBERT embedding than in the cases of word2vec and Law2vec embeddings. It suggests that CNN trained and utilised with this embedding tends to take into account a relatively larger chunk of input sentence while making prediction. 
	\item Described metrics and visualizations provide a peek into the complexity of context handling aspects embedded in a language model. 
	\item It enables an user to identify and catalog attention words in a sentence type for data optimization in downstream processing tasks. 
	
\end{itemize}
 Some issues which need further investigation are:
 \begin{itemize}
 \item Training of these domain-specific models requires time and resources. Apart from algorithmic optimization,  data optimization also plays an important role. Extension of this methodology can be used to remove tokens that do not contribute to the final outcome of any downstream processing tasks. A systematic analysis of the method presented in Section~\ref{section:grad-cam-guided-context-extraction} is warranted.
	\item Mapping of metrics from our methodology to standard machine learning metrics could allow us to infer the quality of language models in a given domain (i.e. legal domain). This allows us to measure the quality of a model when there is not sufficient gold data which can be used for effective training of models (inline to the concept of semi-supervised learning).
	\item An extension of this approach could be used when validating the consistency of context in facts. And inturn the legal argument chain which is built based on these facts. 
 \end{itemize}

\section*{Acknowledgment}
This research was carried out with the support of the Interdisciplinary Centre for Mathematical and Computational Modelling (ICM), University of Warsaw, under grant no GR81-14.

\bibliography{bibliography}
\bibliographystyle{unsrt}

%
%


\end{document}